\documentclass[runningheads]{llncs}
\usepackage[T1]{fontenc}
\usepackage{graphicx}
\usepackage{xurl}
\usepackage{graphicx}
\usepackage[margin=1in]{geometry}
\usepackage{tikz}
\usepackage{pgfplots}
\pgfplotsset{compat=1.18}
\usepackage{adjustbox}
\usepackage{underscore} 
\usepackage{listings}

\usepackage{tabularx}
\usepackage[table]{xcolor}
\definecolor{headerbg}{RGB}{220, 230, 241}   
\definecolor{rowbg}{RGB}{245, 248, 252}      
\definecolor{linecolor}{RGB}{100, 120, 160}  
\usepackage{subcaption}
\usepackage{makecell}

\begin{document}
\title{GROMACS Unplugged: How Power Capping and Frequency Shapes Performance on GPUs}
%
%

\author{Ayesha Afzal\inst{1}\orcidID{0000-0001-5061-0438} \and
Anna Kahler\inst{1} \and
Georg Hager\inst{1}\orcidID{0000-0002-8723-2781} \and
Gerhard Wellein\inst{2}\orcidID{0000-0001-7371-3026}}
\authorrunning{A. Afzal et al.}
%
\institute{Erlangen National High Performance Computing Center (NHR@FAU) \and
Department of Computer Science, Friedrich-Alexander-Universität Erlangen-Nürnberg, Germany\\
\email{\{ayesha.afzal,anna.kahler,georg.hager,gerhard.wellein\}@fau.de}}
\maketitle              
\begin{abstract}
Molecular dynamics simulations are essential tools in computational biophysics, but their performance depend heavily on hardware choices and configuration. In this work, we presents a comprehensive performance analysis of four NVIDIA GPU accelerators -- A40, A100, L4, and L40 -- using six representative \texttt{GROMACS} biomolecular workloads alongside two synthetic benchmarks: Pi Solver (compute bound) and STREAM Triad (memory bound). We investigate how performance scales with GPU graphics clock frequency and how workloads respond to power capping.
The two synthetic benchmarks define the extremes of frequency scaling: Pi Solver shows ideal compute scalability, while STREAM Triad reveals memory bandwidth limits -- framing \texttt{GROMACS}'s performance in context.
Our results reveal distinct frequency scaling behaviors: Smaller \texttt{GROMACS} systems exhibit strong frequency sensitivity, while larger systems saturate quickly, becoming increasingly memory bound. 
Under power capping, performance remains stable until architecture- and workload-specific thresholds are reached, with high-end GPUs like the A100 maintaining near-maximum performance even under reduced power budgets.
Our findings provide practical guidance for selecting GPU hardware and optimizing \texttt{GROMACS} performance for large-scale MD workflows under power constraints.

\keywords{\texttt{GROMACS} \and Molecular dynamics \and GPU benchmarking \and Power cap \and GPU clock frequency}
\end{abstract}

\section{Introduction and related work}

Molecular dynamics (MD) simulations are a cornerstone of computational biophysics, enabling detailed exploration of the structural and dynamical properties of biomolecular benchmarks. By integrating Newton’s equations of motion over time, MD simulations provide atomistic insight into processes such as protein folding, ligand binding, and membrane dynamics~\cite{karplus2002molecular}.
\texttt{GROMACS}~\cite{abraham2015gromacs} is one of the most widely used open-source MD packages due to its algorithmic efficiency, GPU support, and optimization for high-performance computing (HPC) environments. Over recent years, GPU-accelerated simulations have become mainstream in molecular modeling, offering significant speedups and energy efficiency compared to CPU-only execution~\cite{stone2010accelerating,salomon2013routine}.
Modern HPC systems now offer a diverse range of GPU architectures with varying compute capabilities, memory bandwidths, and price points. NVIDIA’s datacenter-class accelerators such as the A100, A40, L4, and L40 represent an evolving hardware landscape for MD practitioners. Selecting optimal hardware and effectively tuning simulations for a given platform remain non-trivial tasks.

In this study, we present a comprehensive benchmarking analysis of \texttt{GROMACS} across four NVIDIA GPU architectures: L4, L40, A40, and A100. Using a curated set of biomolecular benchmarks ranging from small solutes to multi-million-atom viral assemblies, we evaluate simulation throughput (in nanoseconds per day) and investigate the impact of performance tuning strategies -- including GPU clock scaling, power caping -- on overall throughput behavior. Our results provide practical guidance for researchers aiming to select, configure, and optimize MD workloads on heterogeneous GPU-enabled HPC systems.

\paragraph{Related work}

GPU acceleration has long been recognized as transformative for molecular modeling and simulation. Stone et al.~\cite{stone2010accelerating} demonstrated significant speedups in molecular modeling applications using graphics processors, while Salomon-Ferrer et al.~\cite{salomon2013routine} reported routine microsecond-scale MD simulations with AMBER on GPUs, underscoring the role of GPUs in mainstream biomolecular simulation workflows. For \textsc{\texttt{GROMACS}}, Abraham et al.~\cite{abraham2015gromacs} detailed the package’s transition into a highly optimized, multi-level parallel MD engine, with efficient GPU offloading of force calculations and PME electrostatics.  

Beyond algorithmic advances, the energy and performance behavior of GPUs under dynamic hardware tuning has been extensively studied in the HPC community. Early work on GPU DVFS showed that lowering core frequency can substantially reduce power consumption with only modest performance penalties in compute-bound workloads, whereas memory frequency dominates performance in bandwidth-limited applications \cite{hong2010integrated,jia2018characterizing}. Similar results were reported for scientific and machine learning workloads, highlighting the potential of DVFS for energy-efficient GPU computing \cite{song2013energy}.  
Power capping has also emerged as a practical tool for energy-aware HPC system management. Studies have shown that moderate power limits often maintain near-peak performance while reducing energy consumption \cite{haque2017exploiting,ghosh2020power}. While these insights have been applied to general HPC kernels and CPU–GPU systems, only limited work has directly evaluated power capping and frequency scaling in the context of molecular dynamics.  

Our work extends this body of research by systematically combining frequency scaling and power capping analyses on four modern NVIDIA GPU architectures, framed between synthetic compute/memory-bound benchmarks and realistic \texttt{GROMACS} workloads. This dual perspective provides a better
 understanding of how hardware tuning interacts with workload characteristics in MD, offering practical guidance.  

\paragraph{Overview}
This paper is organized as follows: In Sec. \ref{sec:setup}, we describe the benchmark workloads, hardware testbed, and software configuration used in our experiments. Sec. \ref{sec:freq} presents a detailed analysis of GPU frequency scaling, contrasting GROMACS with synthetic compute- and memory-bound workloads. In Sec. \ref{sec:powercap} we investigate the impact of GPU power capping on performance across architectures and workloads. Sec. \ref{sec:conclude} summarizes our findings and outlines directions for future work.

\section{Benchmarks and experimental setup}\label{sec:setup}

This section details the benchmarks, system configurations, and execution environment used for evaluating performance across four modern NVIDIA GPU architectures. The study spans real-world molecular dynamics workloads and synthetic compute/memory-bound microbenchmarks. We further describe clock tuning, power cap enforcement, and software configuration practices used to ensure consistency and reproducibility.

\subsection{Benchmark workloads}

\subsubsection{\texttt{GROMACS} biomolecular benchmarks}
We selected six representative molecular dynamics benchmarks to reflect realistic biophysical workloads with varying sizes and complexities. These include small solvent boxes, protein–membrane systems, and large-scale viral structures, and were sourced from public benchmark repositories. These benchmarks are listed below in order of increasing system size:

\begin{itemize}
\item \textbf{Benchmark 1:} R-143a in hexane (20,248 atoms); a case study from material science domain; characterized by a very high writing-to-file output rate; configured with the outdated Berendsen temperature coupling, and a 1 fs time step.
\item \textbf{Benchmark 2:} Short RNA fragment with explicit water (31,889 atoms); a case study from biochemistry domain; configured with the AMBER14SB force field and OL refinement for RNA and DNA, V-rescale temperature coupling, and a 2 fs time step.
\item \textbf{Benchmark 3:} Protein embedded in membrane with explicit water (80,289 atoms); a case study from life science domain; configured with the outdated Berendsen pressure coupling, V-rescale temperature coupling, a 2 fs time step, and constrained h-bonds.
\item \textbf{Benchmark 4:} Protein in explicit water (170,320 atoms); a case study from biochemistry domain; often used as default benchmark; configured with V-rescale temperature coupling, and a 2 fs time step.
\item \textbf{Benchmark 5:} Protein membrane channel with explicit water (615,924 atoms); configured with V-rescale temperature coupling, Parrinello-Rahman pressure coupling, a 2 fs time step, and constrained h-bonds. Available on GitLab~\url{https://github.com/PDC-support/benchmarks-procurement/tree/master/benchmarks/GROMACS}.
\item \textbf{Benchmark 6:} Large virus protein (1,066,628 atoms); configured with V-rescale temperature coupling, a 2 fs time step, and constrained h-bonds. Available on Zenodo~\url{https://zenodo.org/records/3893789}.
\end{itemize}

Each benchmark was preprocessed into a binary input file (\texttt{.tpr}) using the \texttt{grompp} tool. All benchmarks were run for 200,000 MD steps to ensure a stable sampling window for performance and power metrics.

\subsubsection{Pi Solver compute-bound benchmark}
The CUDA implementation of ``Pi Solver'' approximates $\pi$ via numerical integration of the function $f(x) = \frac{4}{1 + x^2}$. Each thread computes a subinterval of the integration range, storing partial sums in shared memory to reduce global memory traffic and atomic operation overhead. A block-level reduction aggregates these partial results, and a single thread per block updates the global sum using \texttt{atomicAdd}. The kernel is launched with 4096 blocks $\times$ 512 threads to fully utilize the computational resources of modern GPUs, including the NVIDIA A100, A40, L4, and L40. CUDA events are used for precise kernel timing. 

\subsubsection{BabelStream memory-bound benchmark}
BabelStream is a memory bandwidth benchmark modeled after the classic STREAM benchmark. It is designed to assess sustained memory throughput on CPUs and accelerators such as GPUs. It performs a set of simple vector operations -- \texttt{Copy}, \texttt{Mul}, \texttt{Add}, \texttt{Triad}, and \texttt{Dot} -- that represent typical memory-bound kernels in scientific computing. Among these, we focused on the \texttt{Triad} kernel, which performs three memory accesses per iteration (two reads and one write), mimicking the memory behavior of real-world applications like stencil computations and linear algebra solvers. This benchmark allows us to isolate the impact of memory bandwidth on performance across different GPU architectures.

\subsection{Hardware testbed}
Benchmarks were conducted on four modern NVIDIA GPU architectures: L4, L40, A40, and A100. These devices were hosted on high-performance computing (HPC) infrastructure provided by the Erlangen National High Performance Computing Center (NHR@FAU), specifically the \textit{Alex} production cluster and the \textit{Test Cluster} for experimental evaluation.

The \textit{Alex Cluster}\footnote{\url{https://doc.nhr.fau.de/clusters/alex}} is a production HPC system optimized for large-scale GPGPU workloads, particularly in computational sciences such as molecular dynamics. It comprises 38 nodes equipped with dual AMD EPYC 7713 CPUs, 512\,GB DDR4 memory, and eight NVIDIA A40 GPUs (48\,GB GDDR6, 696\,GB/s bandwidth, 37.4\,TFLOP/s FP32). In addition, 20 nodes are configured with dual EPYC 7713 CPUs, 1\,TB RAM, and eight NVIDIA A100 GPUs (40\,GB HBM2), while another 12 nodes feature similar CPUs and memory but with eight NVIDIA A100 GPUs (80\,GB HBM2). The system supports high-throughput I/O via Lustre file systems backed by local NVMe SSDs and is managed using SLURM for batch scheduling.  

\begin{table}[t]
\centering
\caption{Hardware configuration of nodes in the Test Cluster\footref{testcluster}.}
\label{tab:testcluster}
\rowcolors{2}{rowbg}{white} 
\begin{tabularx}{\textwidth}{l *{3}{>{\centering\arraybackslash}X}}
\hline
\rowcolor{headerbg}
\textbf{Hostname} & \textbf{CPU} & \textbf{RAM} & \textbf{GPU Accelerator} \\
\hline
\texttt{icx32} &
\makecell{2× Intel Xeon Platinum 8358\\(32 cores @ 2.6\,GHz)} &
256\,GiB &
\makecell{NVIDIA L4\\(24\,GB GDDR6)} \\
\texttt{genoa2} &
\makecell{2× AMD EPYC 9354\\(32 cores @ 3.25\,GHz)} &
768\,GiB &
\makecell{NVIDIA L40, L40S\\(48\,GB GDDR6)} \\
\hline
\end{tabularx}
\end{table}

The \textit{Test Cluster}\footnote{\label{testcluster}\url{https://doc.nhr.fau.de/clusters/testcluster}} is dedicated to microarchitectural benchmarking and early-access GPU evaluation. It hosts the NVIDIA L4 and L40 GPUs and provides fine-grained control over system-level settings such as performance counters, NUMA placement, and frequency tuning. A summary of the evaluation nodes is shown in Table~\ref{tab:testcluster}. The L4 (icx32) node integrates a low-power GPU optimized for inference and lightweight compute workloads in an Intel Ice Lake-based system, while the L40 (genoa2) node features a high-performance Ada Lovelace-based accelerator with large GDDR6 memory bandwidth, targeted at both graphics and general-purpose compute workloads on a Genoa CPU platform.

\subsection{GPU clock and power management}
To ensure consistent and controlled performance measurements, GPU core and memory clocks were fixed.
To assess performance across different clock settings, exhaustive frequency sweeps were conducted on four NVIDIA GPU architectures: A40, A100, L4, and L40. For each device, custom shell scripts iterated over predefined memory and graphics clock pairs using \texttt{nvidia-smi --\kern0ptapplications-clocks=<MEM\_CLOCK>, \allowbreak<GRAPHICS\_CLOCK>}, and launched jobs via SLURM.
The \texttt{Pi Solver} and \texttt{BabelSTREAM} benchmarks were compiled using \texttt{nvcc} with architecture-specific flags such as \texttt{-arch=sm\_80}, \texttt{sm\_86}, or \texttt{sm\_89}, which target the real (binary) GPU architectures of the A100, A40, and L4/L40 respectively. These flags ensure that the generated code is optimized for the specific compute capabilities of each GPU.
Memory clocks were fixed (A40: 7251\,MHz, A100: 1215\,MHz, L4: 6251\,MHz, L40: 9001\,MHz) while sweeping graphics clock frequencies. Execution time was measured with CUDA events, and outputs were logged per configuration to analyze floating-point throughput sensitivity to frequency scaling.
 
The benchmark was also executed under varying power cap settings using \texttt{nvidia-smi --\kern0pt-power-limit=<W>}. Each run was a dedicated SLURM job on a reserved node with exclusive GPU access (\texttt{CUDA\_VISIBLE\_DEVICES=0}). Power logging was performed using \texttt{nvidia-smi} at 100\,ms intervals, capturing timestamped power draw and GPU utilization. All benchmark outputs, including runtime, power logs, and computed values, were saved in uniquely named directories labeled by GPU model, job ID, and clock or power cap. This setup enables reproducible, large-scale analysis of the performance and efficiency characteristics across modern NVIDIA GPUs under varied operating conditions.


\subsection{Software configuration}
\subsubsection{\texttt{GROMACS}}
All \texttt{GROMACS} simulations were conducted using version 2024.4, compiled with GCC 11.2 and linked against Intel MKL and CUDA libraries for GPU acceleration. Each simulation was executed with one MPI rank per GPU and 16 OpenMP threads (\texttt{-ntmpi 1 -ntomp 16}), ensuring consistent thread and process placement across all architectures. CPU and GPU affinity were explicitly controlled using the \texttt{-pin on} and \texttt{-pinstride 1} options to enhance thread locality and avoid resource contention. GPU acceleration was enabled for all major computational kernels including nonbonded force calculations, PME (Particle-Mesh Ewald) electrostatics, bonded interactions, and integration updates via the options \texttt{-nb gpu}, \texttt{-pme gpu}, \texttt{-bonded gpu}, and \texttt{-update gpu}. Neighbor searching was performed every 20 steps, and each simulation was constrained to a maximum runtime of 0.2 hours using \texttt{-maxh 0.2}. All simulations were performed in single-precision mode using the standard \texttt{gmx mdrun} binary. 

Error correction code (ECC) memory was enabled on both A40 and A100 GPUs to ensure memory reliability during compute-intensive runs. MPI communication was limited to one rank per GPU to reduce inter-process communication overhead. To further improve GPU utilization and minimize measurement noise in power traces, several \texttt{GROMACS}-specific environment variables were configured: \texttt{GMX\_GPU\_PME\_DECOMPOSITION=1} enabled PME-specific load balancing on GPU; \texttt{GMX\_USE\_GPU\_BUFFER\_OPS=1} facilitated direct GPU-side buffer operations; \texttt{GMX\_DISABLE\_GPU\_TIMING=1} disabled internal GPU timing instrumentation; and \texttt{GMX\_ENABLE\_DIRECT\_GPU\_COMM=1} enabled peer-to-peer GPU communication. Collectively, these settings ensured efficient CPU–GPU overlap, reduced kernel launch overheads, and enhanced measurement reproducibility.

\subsubsection{Pi Solver compute-bound benchmark}
The Pi Solver benchmark was compiled using \texttt{nvcc} with archi\-tecture-specific flags (\texttt{sm\_XX}) and \texttt{-O3} performance optimization flag. The solver executed $10^{12}$ integration steps per run using 4096 blocks and 512 threads per block, ensuring a compute-intensive workload suitable for performance evaluation. 

\subsubsection{BabelStream memory-bound benchmark}
We used the BabelStream benchmark, which was cloned from the official GitHub repository\footnote{\url{https://github.com/UoB-HPC/BabelStream}} and compiled with CUDA support. 
The benchmark was built using CMake with architecture-specific flags and the NVIDIA CUDA toolkit on four GPU architectures, using the following command: \texttt{cmake .. -DMODEL=cuda -DCUDA\_ARCH=sm\_XX -DCMAKE\_BUILD\_TYPE=Release -DCMAKE\_CUDA\_COMPILER=\$(which nvcc)}.
Each build was compiled with appropriate \texttt{-arch=sm\_XX} flags: \texttt{sm\_80} for A100, \texttt{sm\_86} for A40, and \texttt{sm\_89} for L4 and L40.
The benchmark was executed using a fixed problem size of 1.5 billion elements (\texttt{ARRAYSIZE = 150994944}) with 10{,}000 iterations (\texttt{NTIMES = 10000}) per run. This configuration ensures a memory-intensive workload that exceeds GPU cache sizes, thereby exercising global memory bandwidth.
In this study, we report only the \texttt{Triad} kernel's peak memory bandwidth (in~{GB/s}) as a representative metric. Other kernels such as \texttt{Copy}, \texttt{Mul}, and \texttt{Add} were omitted from analysis for simplicity.
Power measurements and GPU isolation strategies were identical to those used in the Pi Solver and \texttt{GROMACS} benchmark.

\section{Performance-frequency analysis}\label{sec:freq}
This section analyzes how application performance scales with GPU graphics clock frequency across multiple benchmarks and architectures, highlighting the distinct frequency sensitivity of real-world molecular dynamics workloads (\texttt{GROMACS}) and synthetic compute- and memory-bound kernels (Pi Solver and STREAM Triad).

\begin{table}[t]
\centering
\caption{Configurable memory and graphics clock frequency ranges for NVIDIA GPUs.}
\label{tab:freq_range}
\rowcolors{2}{rowbg}{white} 
\begin{tabularx}{0.8\textwidth}{l *{2}{>{\centering\arraybackslash}X}}
\hline
\rowcolor{headerbg}
\textbf{NVIDIA GPUs} & \textbf{Memory clock} & \textbf{Graphics clock range} \\
\rowcolor{headerbg} & \textbf{[MHz]} & \textbf{[MHz]} \\
\hline
A40  & 405 or 7251  & 210 -- 1740 (step 15) \\
A100 & 1215         & 210 -- 1410 (step 15) \\
L4   & 405 or 6251  & 210 -- 2040 (step 15) \\
L40  & 405 or 9001  & 210 -- 2490  (step 15) \\
\hline
\end{tabularx}
\end{table}

Table~\ref{tab:freq_range} summarizes the range of configurable GPU graphics clock frequencies across NVIDIA datacenter GPUs. Frequency settings were swept over extensive ranges to analyze performance sensitivity. Each frequency pair specifies the memory and graphics clock frequencies, with memory frequencies kept mostly constant while graphics frequencies varied widely. These settings enable detailed characterization of the frequency-performance relationship under realistic and extreme conditions. Frequency ranges differ across GPUs due to architectural and BIOS constraints, with minimum and maximum values reflecting stable operational limits.

All experiments shown in this work used standard memory frequencies. Although a lower 405~MHz memory mode (the idle or P8 power state) is available on three tested GPUs, it was excluded from performance analysis due to its impractically low bandwidth, which does not reflect realistic application behavior. 
We observe that high-throughput GPUs like the A40, and L40 can hit bandwidth bottlenecks when graphics clocks exceed memory clocks fixed at 405~MHz, starving streaming multiprocessors (SMs) for data in memory-bound workloads. However, the L4, due to its lower SM throughput and narrower memory bus, is less sensitive to memory underclocking.



\begin{figure}[htbp]
    \centering
    
    \subfloat[Benchmark 1\label{fig:fl_md1_berendsen}]{
        \includegraphics[width=0.45\textwidth]{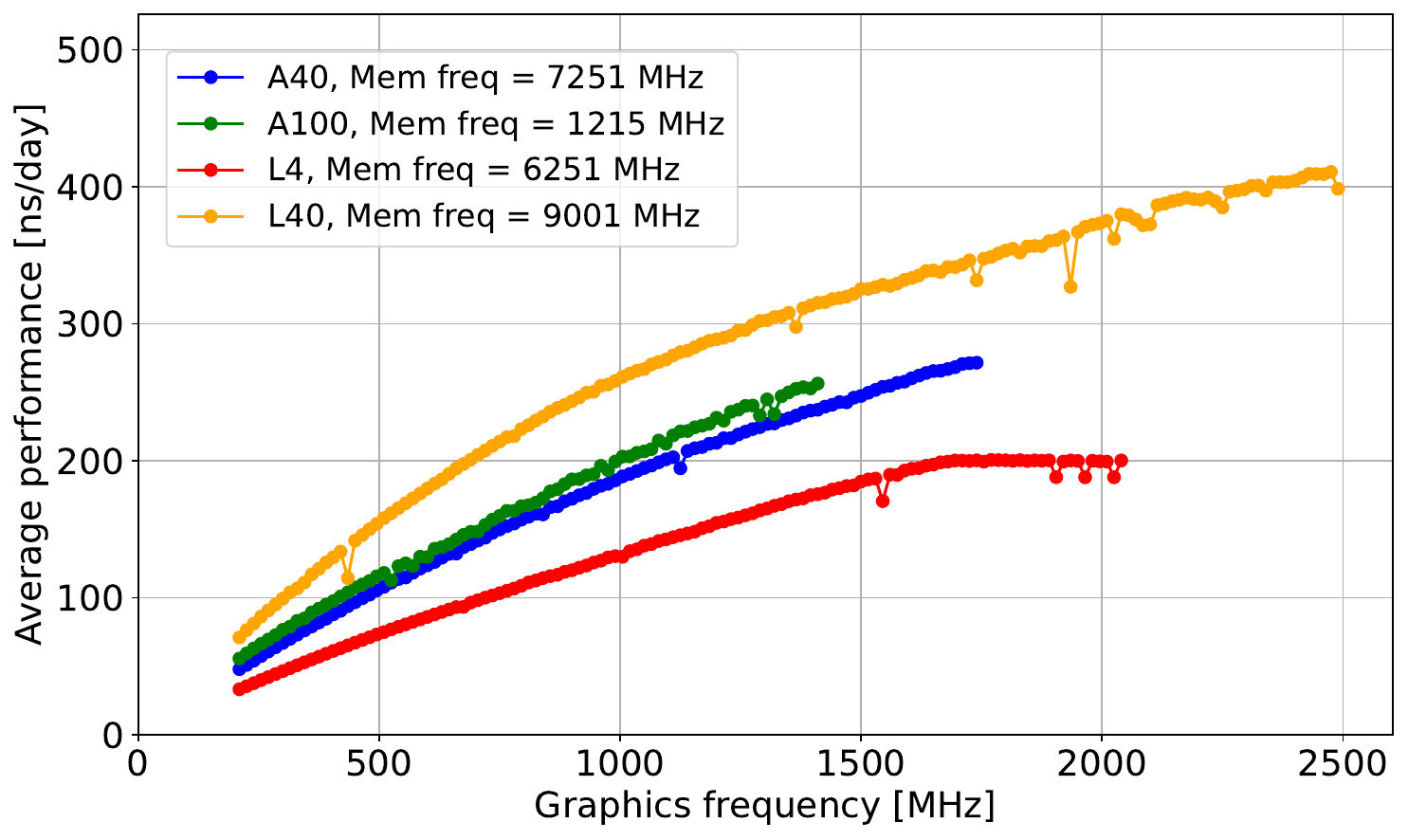}
    }
    \hfill
    \subfloat[Benchmark 2\label{fig:rnanvt}]{
        \includegraphics[width=0.45\textwidth]{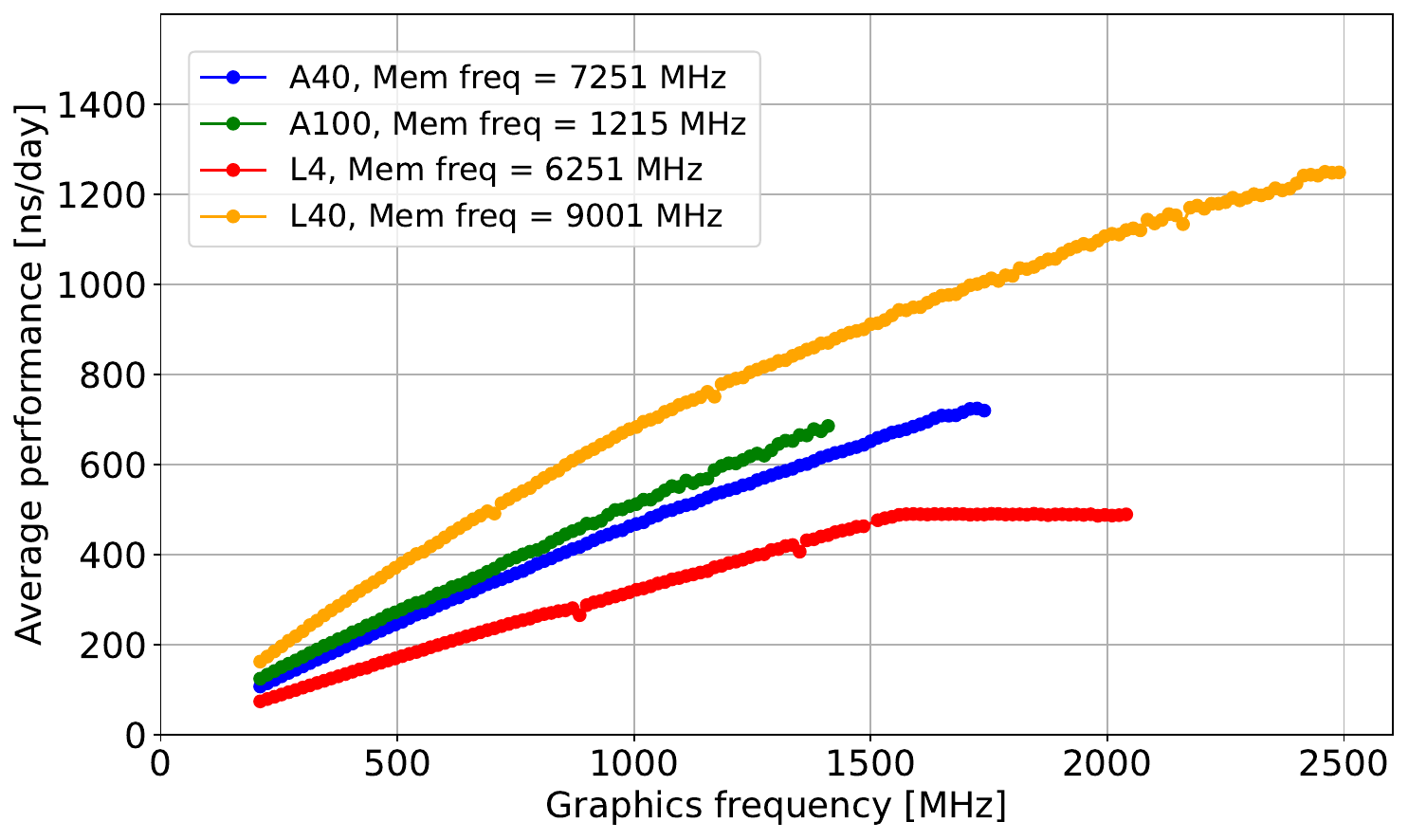}
    }

    \vspace{0.5cm}
    \subfloat[Benchmark 3\label{fig:2md_start0}]{
        \includegraphics[width=0.45\textwidth]{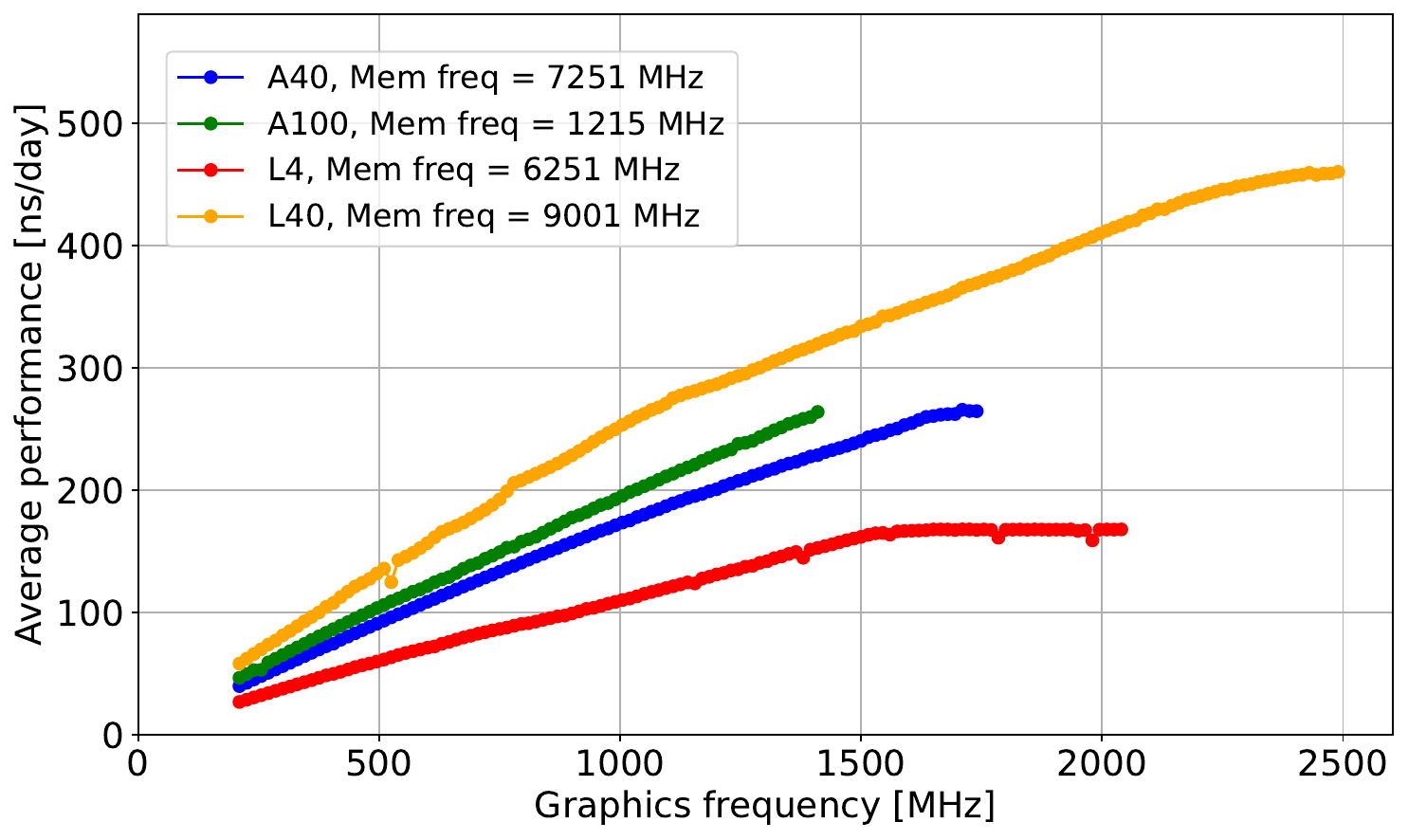}
    }
    \hfill
    \subfloat[Benchmark 4\label{fig:pi_large_test}]{
        \includegraphics[width=0.45\textwidth]{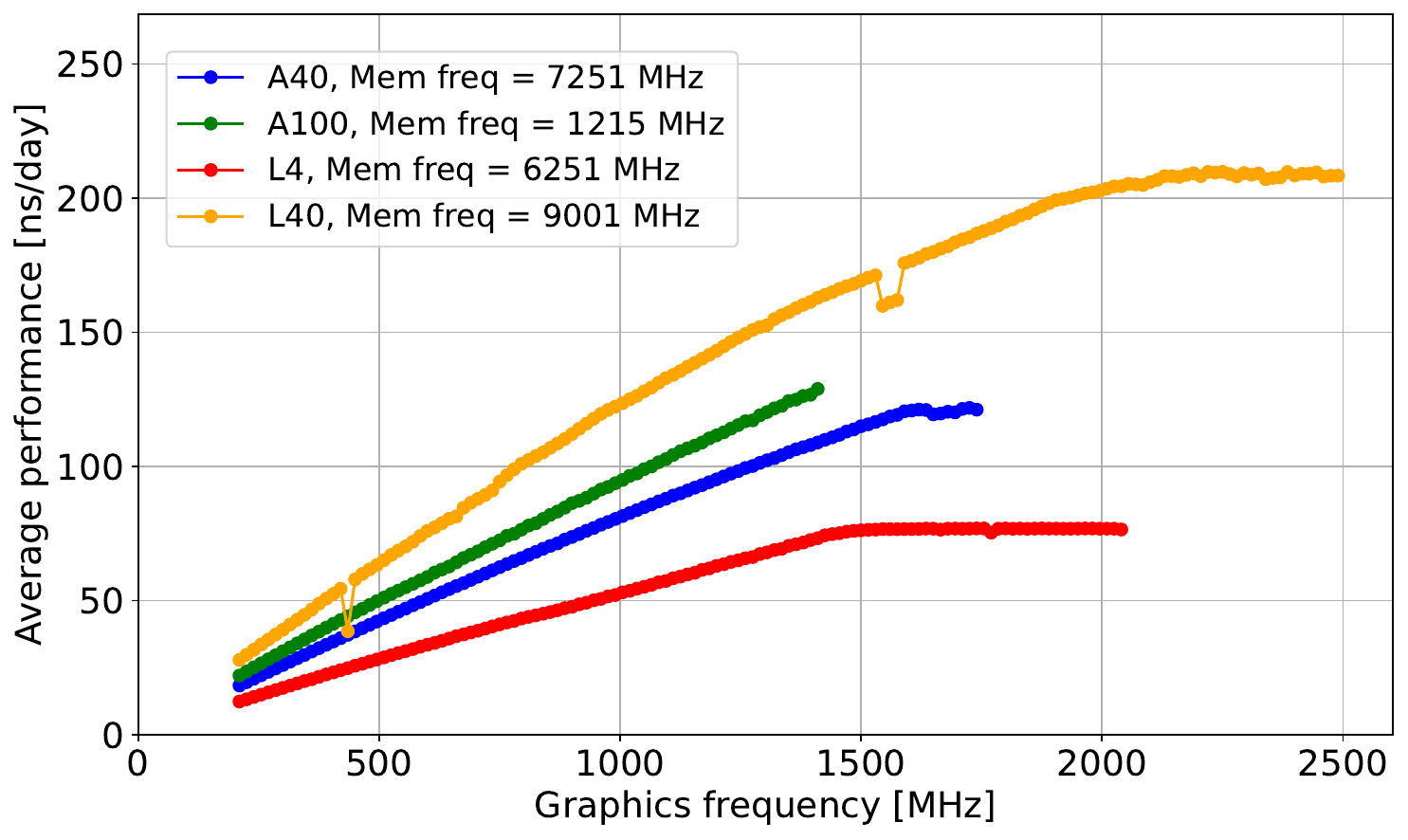}
    }

    \vspace{0.5cm}
    
    \subfloat[Benchmark 5\label{fig:eag1}]{
        \includegraphics[width=0.45\textwidth]{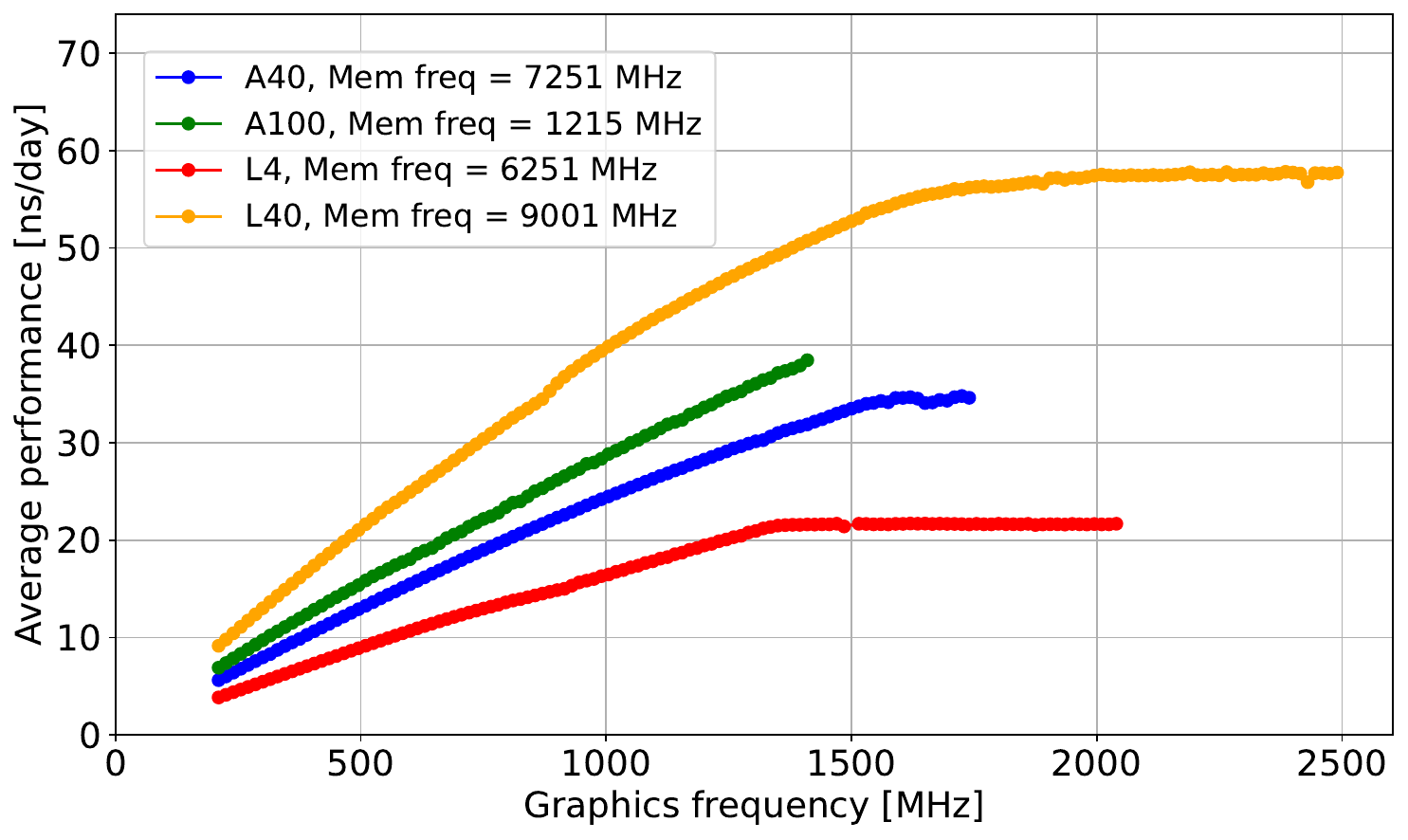}
    }
    \hfill
    \subfloat[Benchmark 6\label{fig:stmv_pme_nvt}]{
        \includegraphics[width=0.45\textwidth]{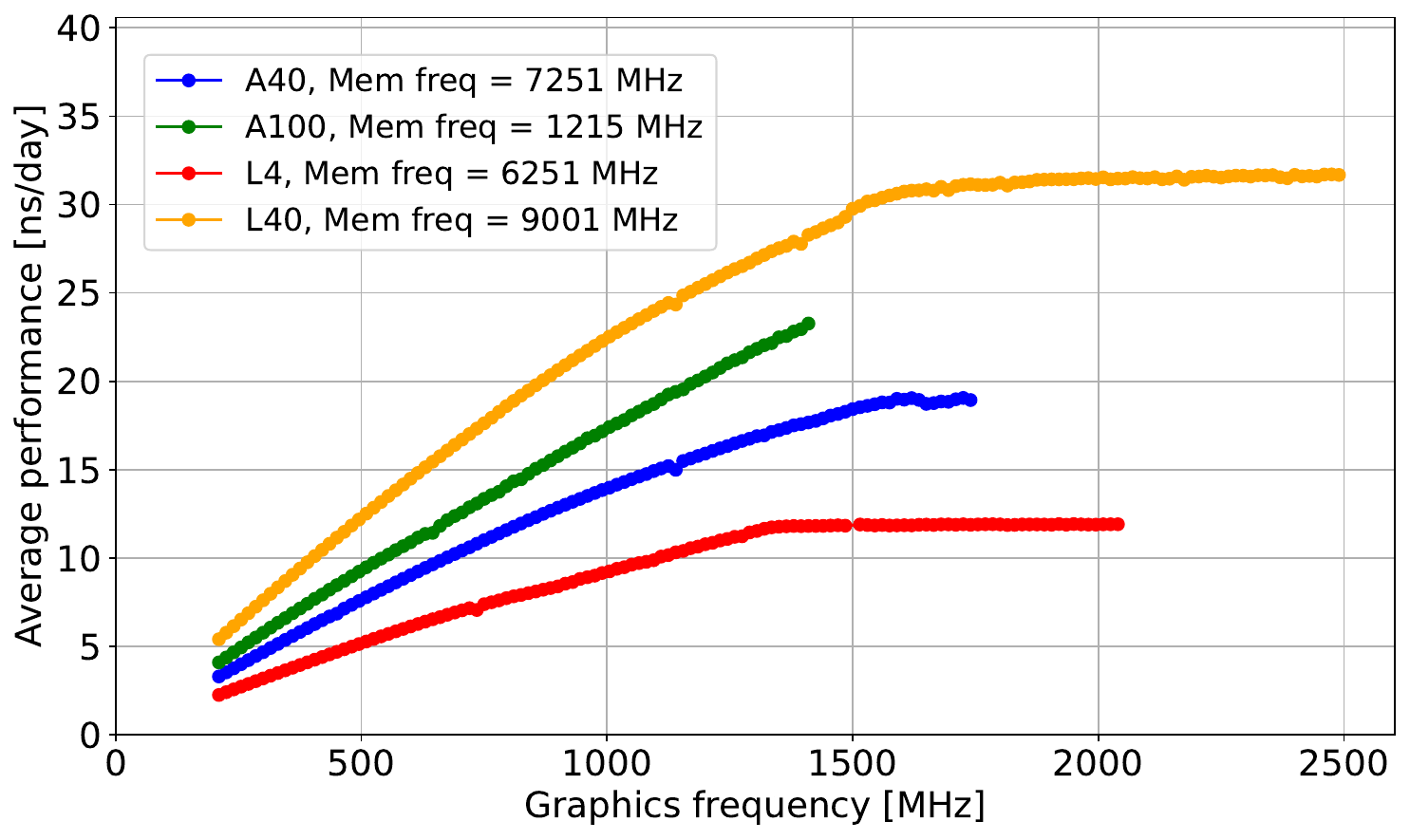}
    }
    
    \caption{Average performance (ns/day) as a function of GPU graphics clock frequency, measured at the maximum memory frequency setting, for various accelerators across six GROMACS benchmarks.}
    \label{fig:Combined_perf_vs_freq}
\end{figure}

\subsection{\texttt{GROMACS}}
Figure~\ref{fig:Combined_perf_vs_freq} compares the average throughput (in ns/day) across six \texttt{GROMACS} biomolecular benchmarks on the A40, A100, L4, and L40 GPUs, plotted as a function of GPU graphics clock frequency for each GPU type. Benchmark size and complexity have a significant impact on the observed scaling behavior. As expected, with increasing GPU frequency the smallest benchmarks (1 and 2) exhibit steep performance scaling and the medium-scale benchmarks (3 and 4) show more gradual scaling and earlier saturation. In contrast,the largest benchmark casess (5 and 6) show reduced frequency sensitivity with flat curves at higher frequencies, indicating they might be more memory bound and limited by data movement rather than compute throughput.
This behavior highlights the increasing role of memory bandwidth and interconnect limitations as benchmark size grows, where raising the graphics frequency provides little additional benefit.
Overall, the benchmark analysis demonstrates a clear transition: small systems are frequency-sensitive, medium systems exhibit balanced behavior, and large systems have negligible sensitivity to frequency tuning. 

Across all six biomolecular benchmarks, distinct patterns emerge in how GPU architectures respond to graphics frequency scaling. The L40 consistently outperformed other tested GPUs across all benchmark sizes by a significant margin, highlighting its superior memory bandwidth and computational scaling especially for large, communication-intensive MD workloads. The A40 and A100 deliver intermediate performance levels, with the A40 generally closer to the A100 on smaller benchmarks, while the A100 pulls ahead on larger cases. This suggests that architectural differences such as clock frequency limits and memory hierarchy can impact smaller benchmarks differently. In contrast, the low-power L4 GPU exhibits the lowest overall throughput across all benchmarks and shows clear limitations on the largest cases. However, it still achieves reasonable performance on smaller benchmarks at modest clock frequencies, making it a potential candidate for lightweight MD workloads. Its scaling benefits saturate quickly, emphasizing its suitability for energy-efficient or edge-compute MD applications.

The high-end A100 GPU shows no performance saturation across benchmarks and is largely insensitive to graphics frequency changes due to ample compute headroom and memory bandwidth. Conversely, smaller accelerators like the L4 rely heavily on maintaining higher frequencies to sustain throughput. Mid-range GPUs (A40, L40) display intermediate characteristics: they benefit from frequency scaling on smaller workloads but converge toward similar performance levels as workloads increase in size and complexity. For example, increasing the graphics frequency from 1.5~GHz to 2.5~GHz on large benchmarks yields no performance gain on the L40, implying memory bandwidth is fully utilized.

\begin{figure}[t]
    \centering
    \begin{subfigure}[t]{\textwidth}
        \centering
        \begin{tikzpicture}
        \begin{axis}[
            ybar,
            symbolic x coords={Benchmark 1, Benchmark 2, Benchmark 3, Benchmark 4, Benchmark 5, Benchmark 6},
            xtick=data,
            ylabel={Performance (ns/day)},
            bar width=12pt,
            width=\textwidth,
            height=0.32\textwidth,
            enlarge x limits=0.09,
            nodes near coords,
            nodes near coords align={vertical},
            nodes near coords=\rotatebox{90}{\pgfmathprintnumber\pgfplotspointmeta},
            ymin=0,
            ymax=1700,
            legend style={
                at={(0.75,0.95)},
                anchor=north east,
                legend columns=4,
                draw=none,
                fill=none,
                font=\small
            }
        ]
        \addplot coordinates {(Benchmark 1, 272) (Benchmark 2, 720) (Benchmark 3, 264)(Benchmark 4, 121)  (Benchmark 5, 35) (Benchmark 6, 19)}; 
        \addplot coordinates {(Benchmark 1, 256.385) (Benchmark 2, 685.762)(Benchmark 3, 264.115)(Benchmark 4, 128.896)  (Benchmark 5, 38.476)  (Benchmark 6, 23.268)}; 
        \addplot coordinates {(Benchmark 1, 200)(Benchmark 2, 489) (Benchmark 3, 168)  (Benchmark 4, 77) (Benchmark 5, 22) (Benchmark 6, 12)};  
        \addplot coordinates {(Benchmark 1, 399) (Benchmark 2, 1249) (Benchmark 3, 460)(Benchmark 4, 208)  (Benchmark 5, 58) (Benchmark 6, 32)}; 
        \legend{A40, A100, L4, L40}
        \end{axis}
        \end{tikzpicture}
        \caption{Maximum GPU graphics frequencies: 1.74 GHz (A40), 1.41 GHz (A100), 2.04 GHz (L4), 2.49 GHz (L40)}        
        \label{fig:gpu_comparison_a}
    \end{subfigure}

    \vspace{1em}

    \begin{subfigure}[t]{\textwidth}
        \centering
        \begin{tikzpicture}
        \begin{axis}[
            ybar,
            symbolic x coords={Benchmark 1, Benchmark 2, Benchmark 3, Benchmark 4, Benchmark 5, Benchmark 6},
            xtick=data,
            ylabel={Performance (ns/day)},
            bar width=12pt,
            width=\textwidth,
            height=0.32\textwidth,
            enlarge x limits=0.09,
            nodes near coords,
            nodes near coords align={vertical},
            nodes near coords=\rotatebox{90}{\pgfmathprintnumber\pgfplotspointmeta},
            ymin=0,
            ymax=300,
            legend style={
                at={(0.75,0.95)},
                anchor=north east,
                legend columns=4,
                draw=none,
                fill=none,
                font=\small
            }
        ]
\addplot coordinates {
    (Benchmark 1, 47.9445)
    (Benchmark 2, 107.421)
    (Benchmark 3, 39.8595)
    (Benchmark 4, 18.3825)
    (Benchmark 5, 5.6175)
    (Benchmark 6, 3.3005)
}; 

\addplot coordinates {
    (Benchmark 1, 55.648)
    (Benchmark 2, 124.681)
    (Benchmark 3, 46.585)
    (Benchmark 4, 22.099)
    (Benchmark 5, 6.894)
    (Benchmark 6, 4.096)
}; 

\addplot coordinates {
    (Benchmark 1, 33.264)
    (Benchmark 2, 74.3275)
    (Benchmark 3, 26.874)
    (Benchmark 4, 12.4165)
    (Benchmark 5, 3.841)
    (Benchmark 6, 2.253)
}; 

\addplot coordinates {
    (Benchmark 1, 71.07375)
    (Benchmark 2, 162.662)
    (Benchmark 3, 58.2445)
    (Benchmark 4, 27.908)
    (Benchmark 5, 9.1465)
    (Benchmark 6, 5.4085)
}; 

        \legend{A40, A100, L4, L40}
        \end{axis}
        \end{tikzpicture}
        \caption{Minimum GPU graphics frequencies: 0.21 GHz (A40, A100, L4, L40)}
        \label{fig:gpu_comparison_b}
    \end{subfigure}
    \caption{Comparison of average throughput (ns/day) across six biomolecular systems at maximum GPU memory frequency, i.e., 7.251 GHz (A40), 1.215 GHz (A100), 6.251 GHz (L4) and 9.001 GHz (L40).}
    \label{fig:gpu_comparison}
\end{figure}
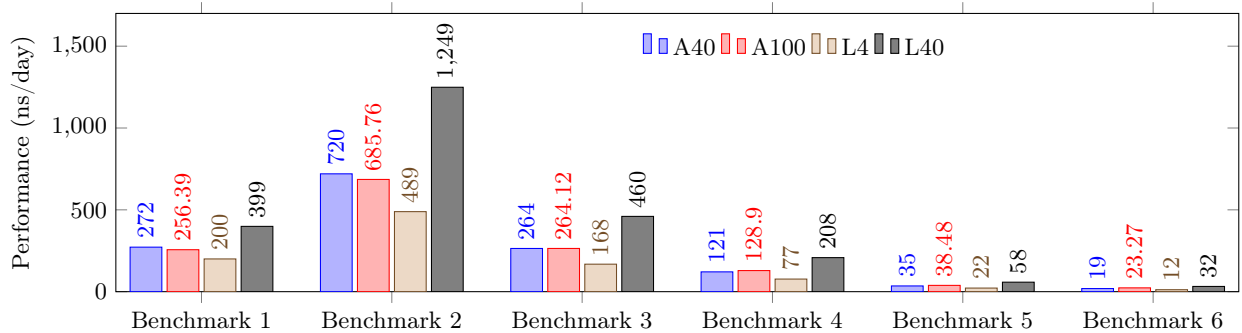
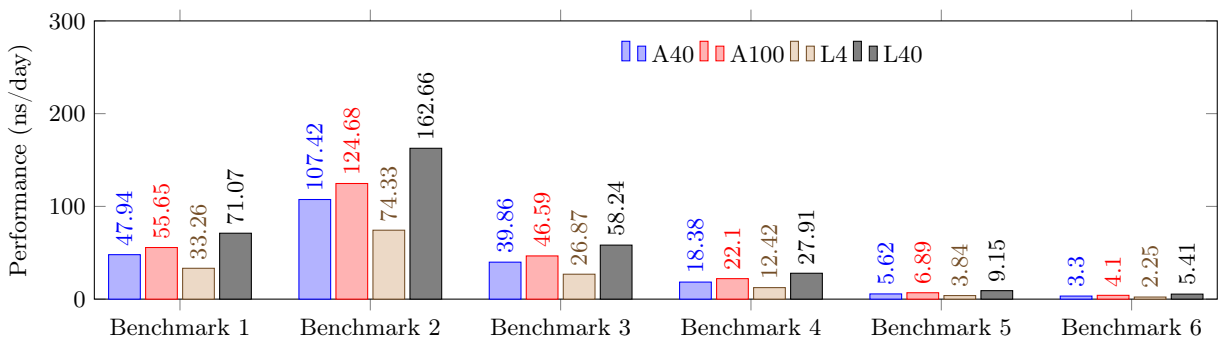

Figure~\ref{fig:gpu_comparison} compares GPU performance at (a) maximum and (b) minimum graphics frequencies with memory clocks fixed at each GPU's peak. At maximum graphics frequency, the L40 leads across all benchmarks, followed by the A100 and A40, with the A40 performing closer to the A100 on smaller cases. The L4 lags overall but performs reasonably on small benchmarks. At the lowest graphics frequency (0.21 GHz), all GPUs show significant performance drops, especially on smaller benchmarks. Larger benchmarks degrade less, confirming limited sensitivity to core frequency in bandwidth-constrained regimes. Notably, the L40 maintains its lead even at low frequency, underscoring architectural efficiency.
These results highlight that high frequencies benefit smaller benchmarks, while larger benchmarks saturate early, limiting further gains.

This progression underscores the importance of selecting both hardware and frequency settings in accordance with benchmark size to maximize efficiency in molecular dynamics simulations.

\begin{figure}[t]
    \centering
    
    \subfloat[Pi Solver\label{fig:STREAM}]{
            \includegraphics[width=0.45\textwidth]{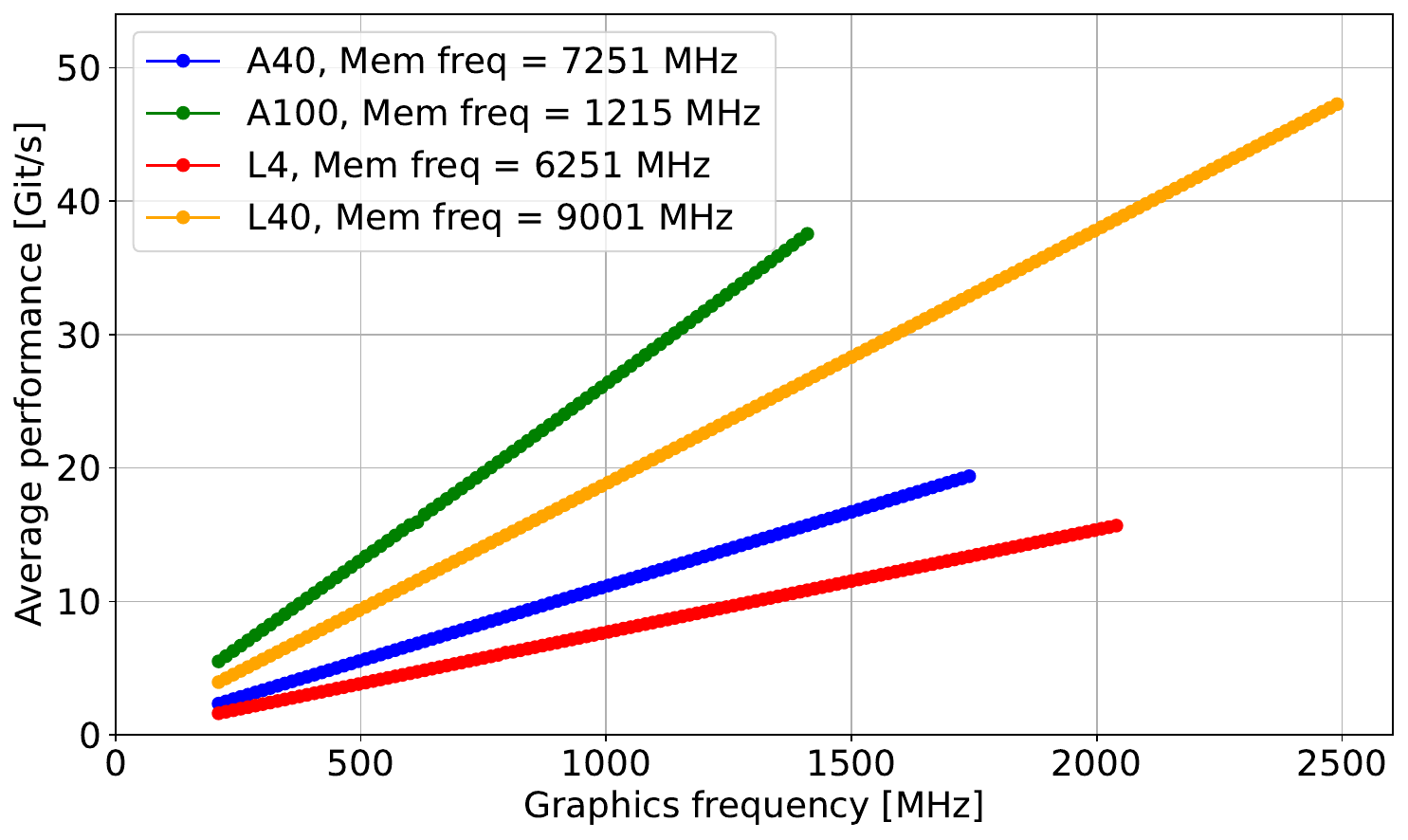}
    }
    \hfill
    \subfloat[STREAM Triad\label{fig:PISOLVER}]{
        \includegraphics[width=0.45\textwidth]{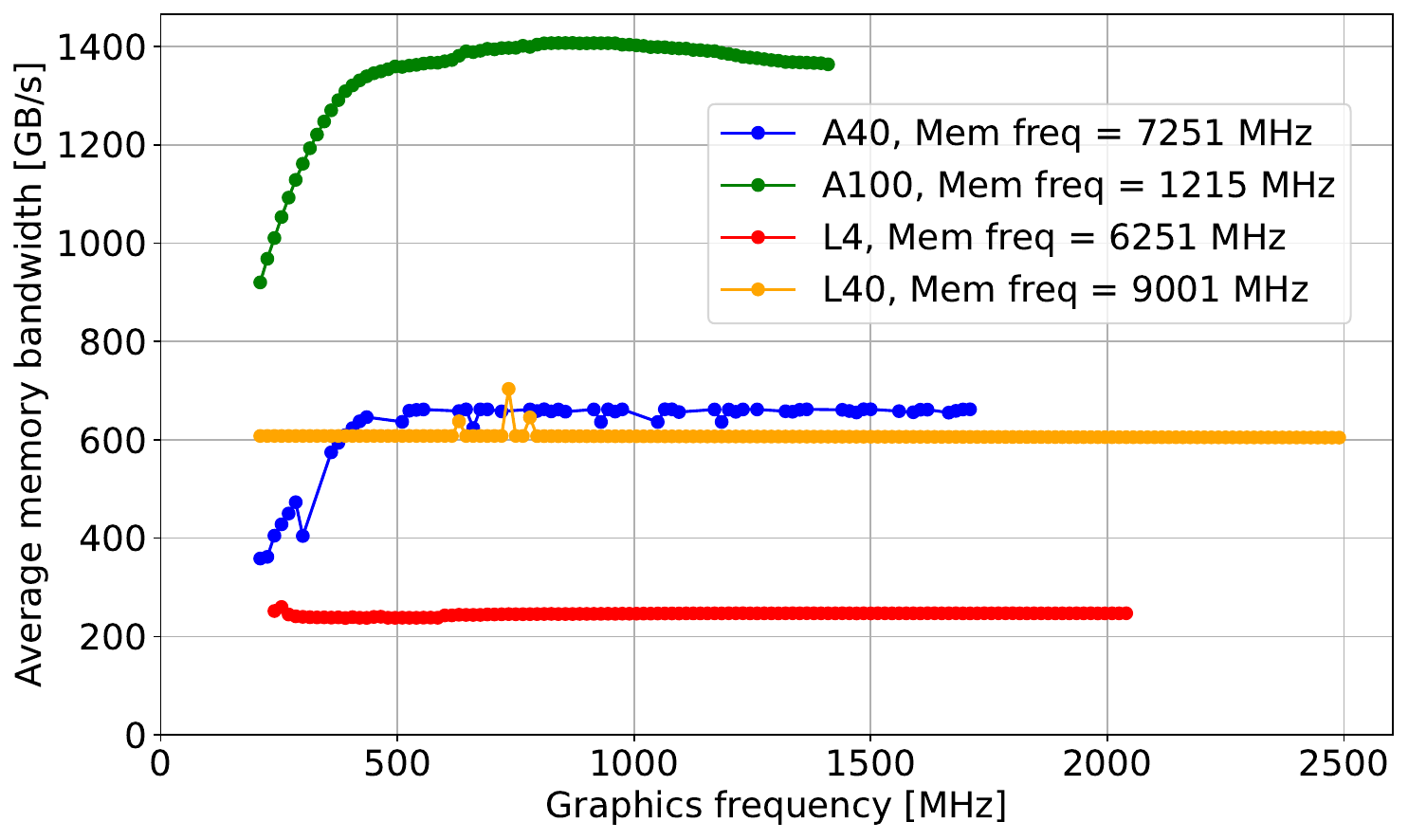}
    }
    
    \caption{Average performance as a function of graphics frequency for various accelerators at the maximum memory frequency setting, across the Pi solver and STREAM TRIAD benchmarks.}
    \label{fig:PisolverSTREAM_perf_vs_freq}
\end{figure}

\subsection{Pi Solver \& STREAM Triad}
Figure~\ref{fig:PisolverSTREAM_perf_vs_freq} presents the average throughput of the (a) Pi Solver and (b) BabelStream Triad benchmarks on the A40, A100, L4, and L40 GPUs, plotted as a function of GPU graphics clock frequency. These benchmarks isolate the impact of graphics frequency on compute- and memory-bound workloads, serving as reference points to assess \texttt{GROMACS}'s sensitivity to frequency and its balance of compute and memory demands.

Pi Solver exhibits linear scaling with graphics frequency across all architectures, consistent with its compute-bound nature and high arithmetic intensity. Its performance improves proportionally with frequency, indicating minimal memory bottlenecks and effective utilization of available compute resources.

In contrast, BabelStream Triad shows a saturating performance trend as frequency increases. As a memory-bound benchmark, its throughput saturates once memory bandwidth is fully utilized—especially on GPUs like the A100 and L40 with high-bandwidth memory systems. Beyond this point, increasing clock speed offers little to no performance gain.

These two benchmarks represent opposite ends of the frequency-scaling spectrum: Pi Solver demonstrates ideal scalability for compute-heavy workloads, while STREAM Triad highlights the limitations imposed by memory bandwidth. Together, they establish performance boundaries that help contextualize \texttt{GROMACS}'s behavior under frequency scaling.

\section{Performance-power capping analysis}\label{sec:powercap}
This section examines the impact of power capping on GPU performance across various benchmarks and architectures, exploring how real-world \texttt{GROMACS} and synthetic compute- and memory-bound Pi Solver and STREAM Triad workloads respond to enforced power limits and identifying their sensitivity to power constraints.

Table~\ref{tab:powercap} summarizes the configurable power capping ranges for four NVIDIA GPUs. Power capping is applied to limit the power consumption of the GPU, which can help with thermal constraints or datacenter power budgeting. \emph{Thermal Design Power (TDP)} refers to the typical power draw under full load without any power limit applied. \emph{Power capping range} indicates the range over which power can be manually limited. The \emph{minimum cap} may vary depending on the system's BIOS or thermal configuration and might not be reachable in all environments. Setting a cap below the hardware-enforced minimum resulted in an error. 

\begin{table}[htb]
\centering
\caption{Configurable power capping ranges and Thermal Design Power (TDP) for NVIDIA GPUs.}
\label{tab:powercap}
\rowcolors{2}{rowbg}{white} 
\begin{tabularx}{0.8\textwidth}{l *{2}{>{\centering\arraybackslash}X}}
\hline
\rowcolor{headerbg}
\textbf{NVIDIA GPUs} & \textbf{TDP} & \textbf{Power capping range} \\
\rowcolor{headerbg} & \textbf{[W]} & \textbf{[W]} \\
\hline
A40     & 300   & 100 -- 300 \\
A100    & 400   & 100 -- 400 \\
L4      & 72    & 40 -- 72   \\
L40     & 300   & 100 -- 300 \\
\hline
\end{tabularx}
\end{table}

\begin{figure}[htbp]
    \centering
    
    \subfloat[Benchmark 1\label{fig:fl_mdBenchmark 11_berendsen}]{
        \includegraphics[width=0.45\textwidth]{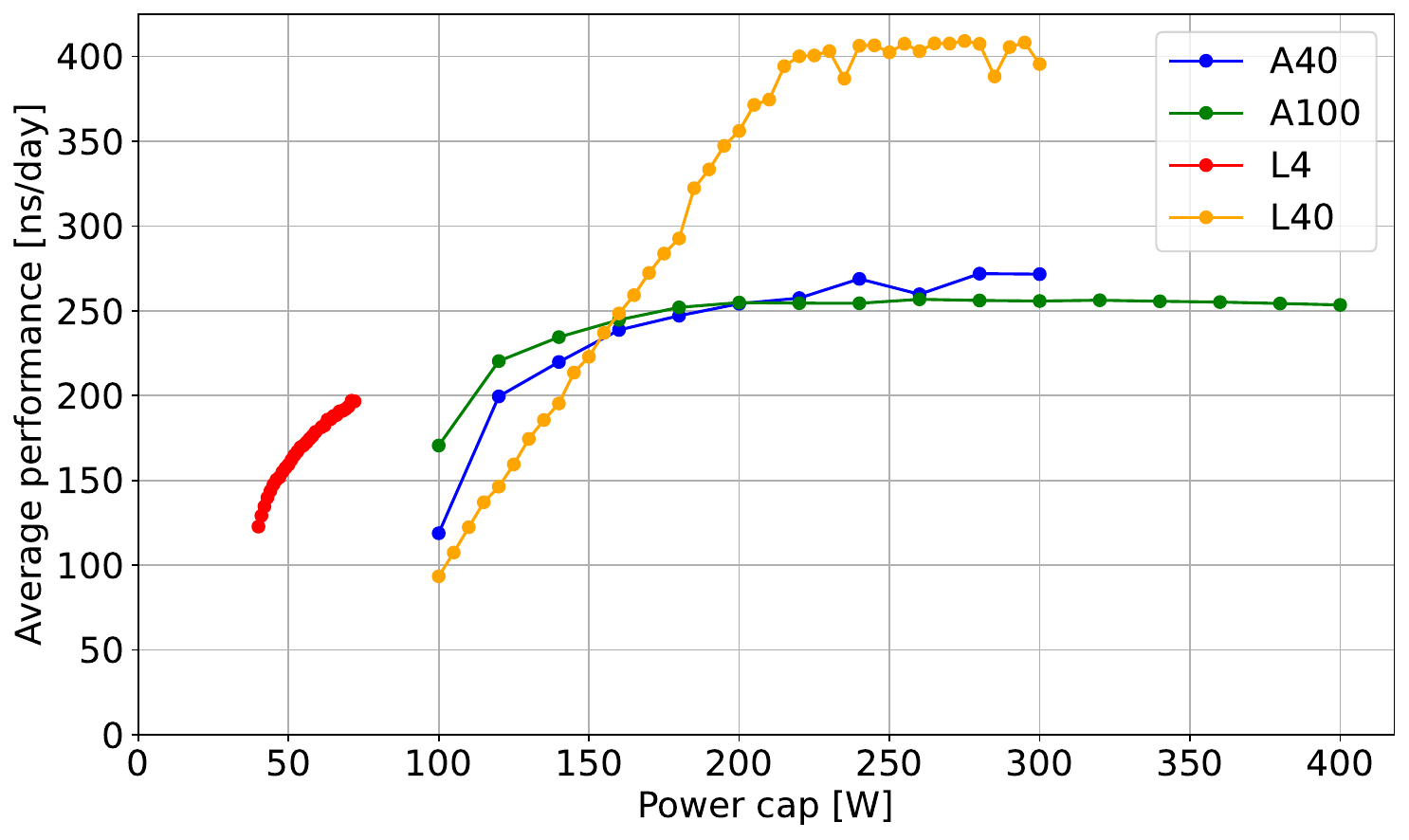}
    }
    \hfill
    \subfloat[Benchmark 2\label{fig:rnanvt}]{
        \includegraphics[width=0.45\textwidth]{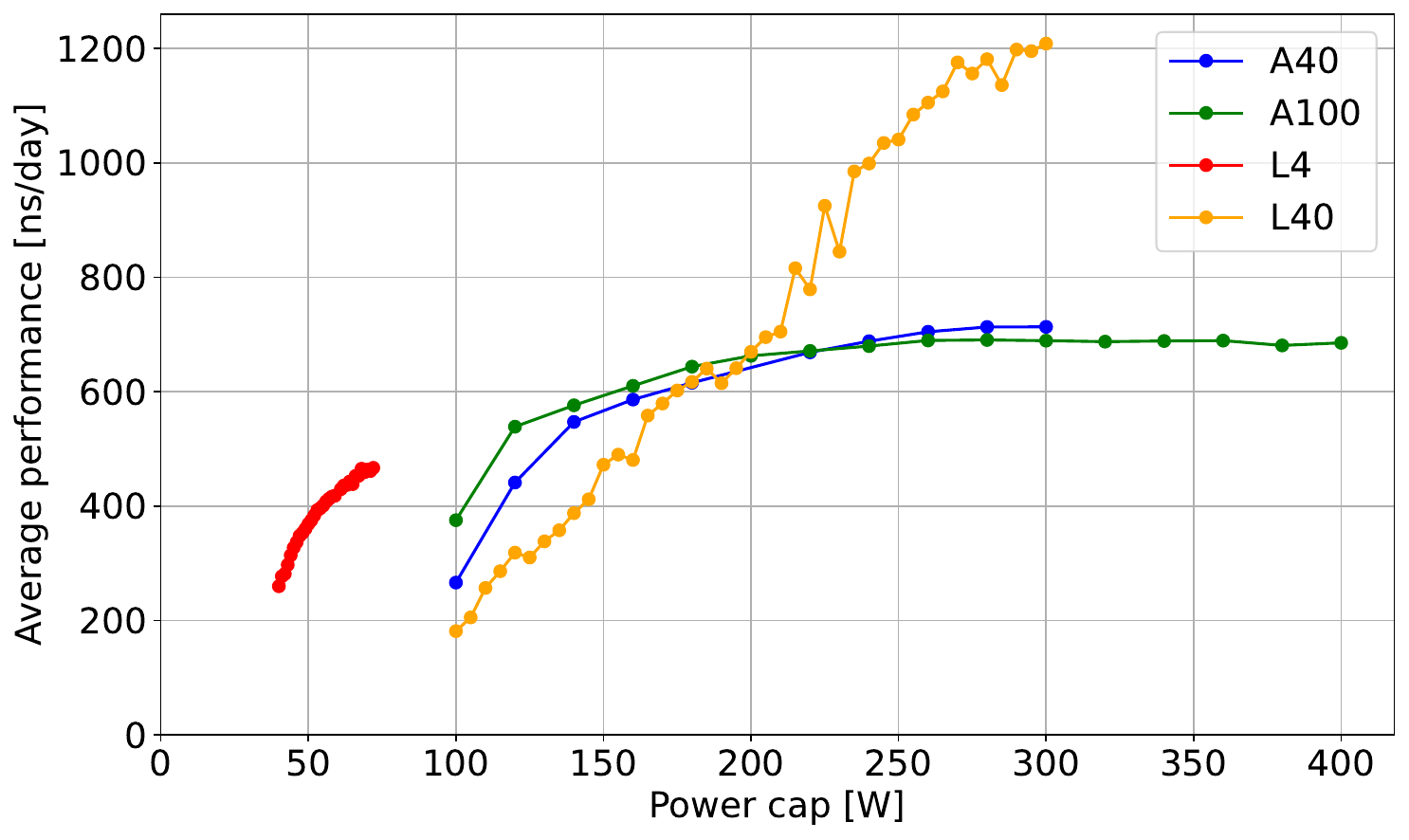}
    }

    \vspace{0.5cm}
    \subfloat[Benchmark 3\label{fig:2md_start0}]{
        \includegraphics[width=0.45\textwidth]{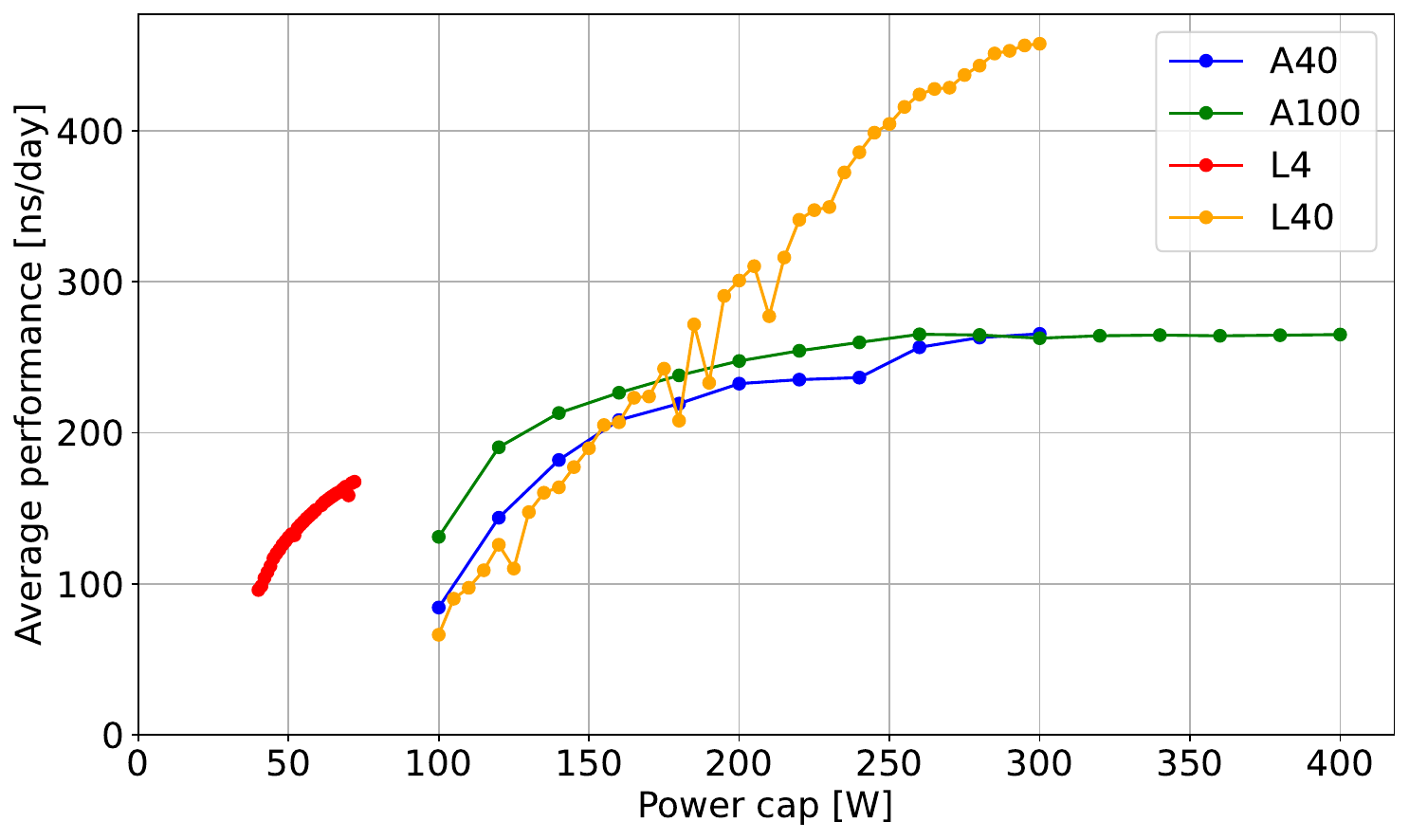}
    }
    \hfill
    \subfloat[Benchmark 4\label{fig:pi_large_test}]{
        \includegraphics[width=0.45\textwidth]{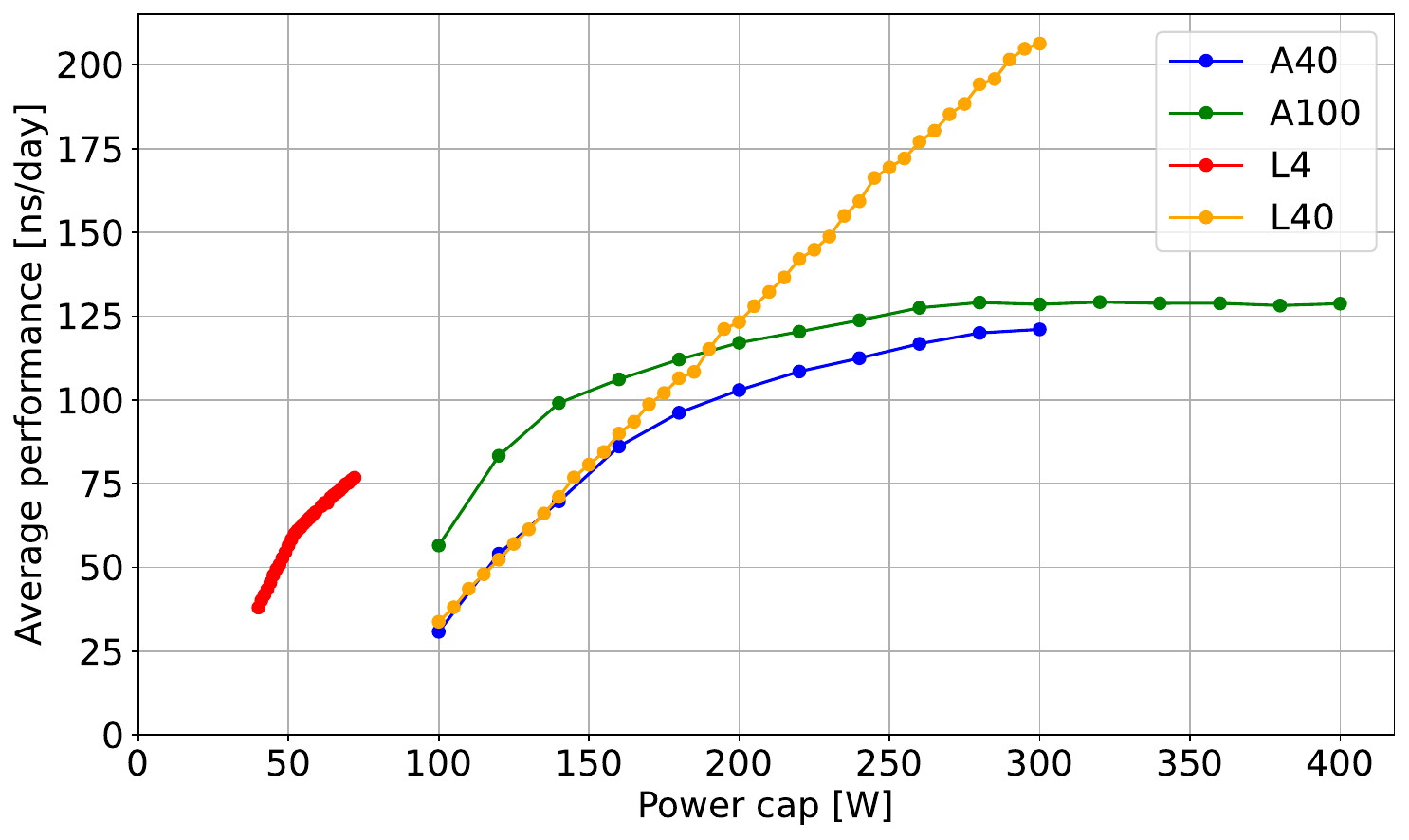}
    }

    \vspace{0.5cm}
    
    \subfloat[Benchmark 5\label{fig:eag1}]{
        \includegraphics[width=0.45\textwidth]{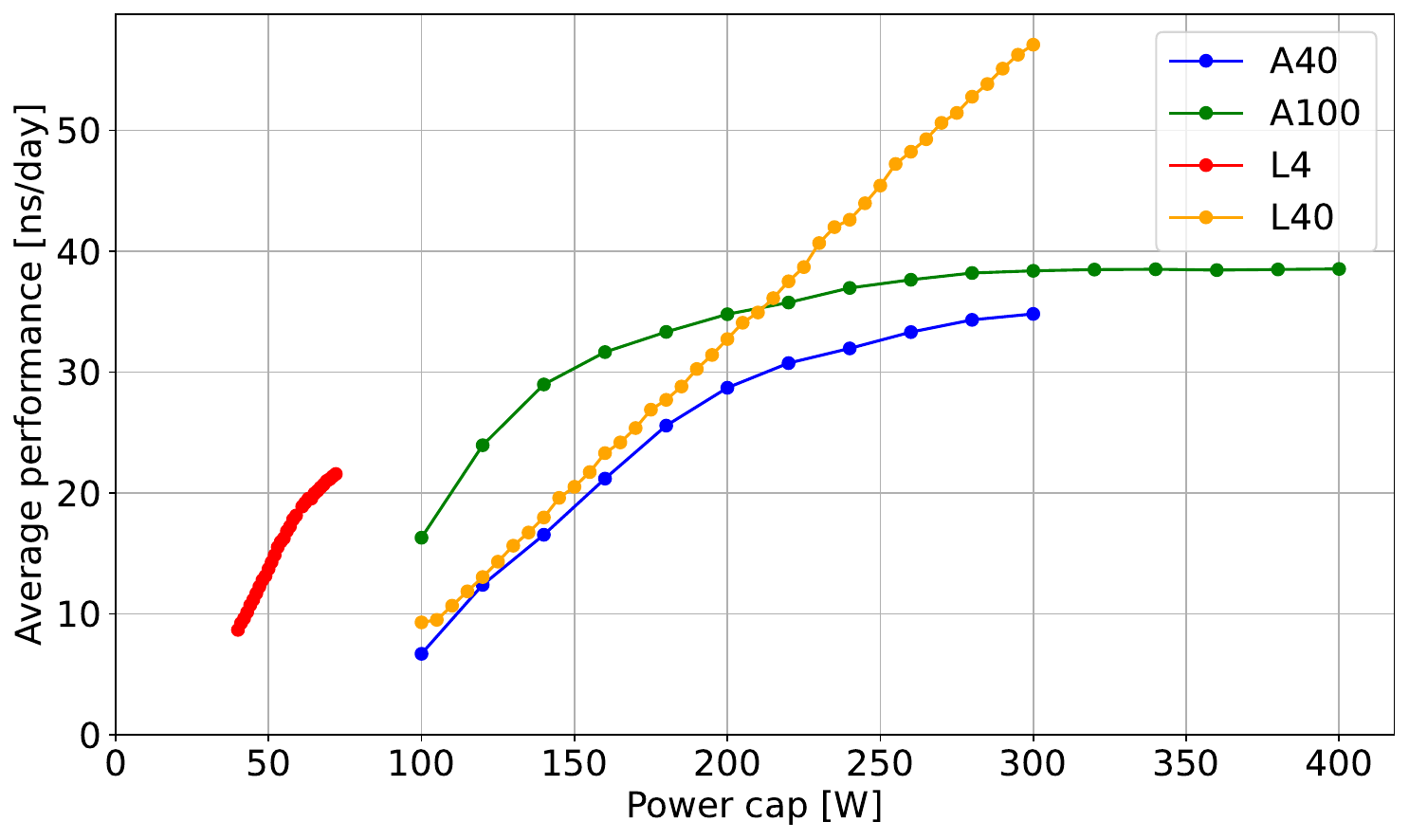}
    }
    \hfill
    \subfloat[Benchmark 6\label{fig:stmv_pme_nvt}]{
        \includegraphics[width=0.45\textwidth]{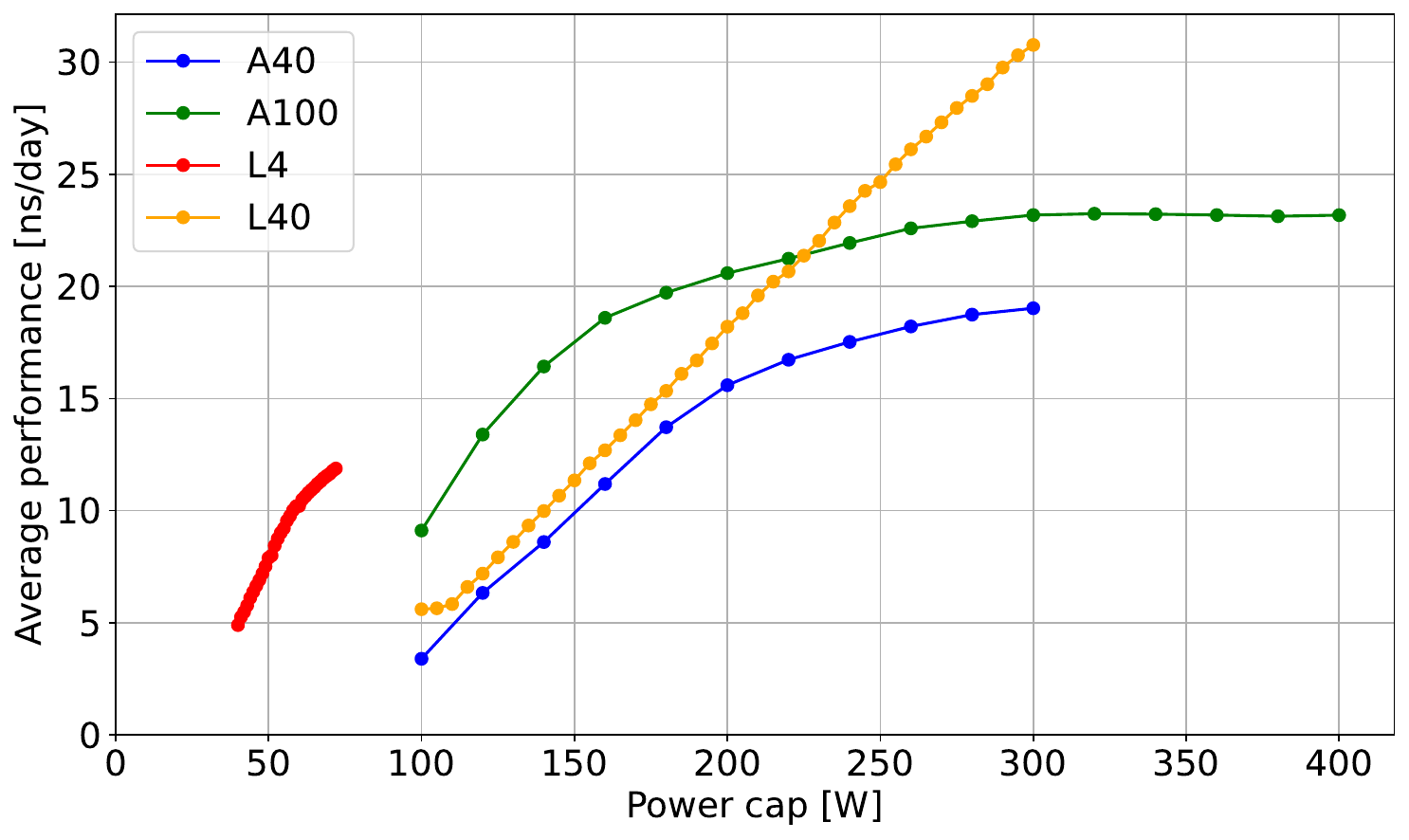}
    }
    
    \caption{Average performance (ns/day) as a function of GPU power cap, measured at the maximum memory frequency setting of each accelerator, for various accelerators across six GROMACS benchmarks.
    }
    \label{fig:Combined2_perf_vs_powercap}
\end{figure}

\subsection{\texttt{GROMACS}}
Figure~\ref{fig:Combined2_perf_vs_powercap} shows the average throughput (in ns/day) across six \texttt{GROMACS} biomolecular benchmarks on the A40, A100, L4, and L40 GPUs as a function of power cap.
Performance remains consistent between the maximum power cap and maximum graphics frequency settings, indicating that both configurations enable similar peak throughput for \texttt{GROMACS} workloads.
We observe some variation in \texttt{gmx mdrun} results (notably with the L40) due to \texttt{GROMACS} dynamically optimizing execution to maximize throughput. Since we did not enable the non-production reproducibility mode (\texttt{mdrun -reprod}), such variation is expected. This reproducibility mode\footnote{\url{https://manual.gromacs.org/documentation/current/user-guide/mdrun-features.html#running-a-simulation-in-reproducible-mode}} ensures binary-identical results on the same hardware/software stack but is not portable across systems or compilers. It is typically used for debugging rather than benchmarking.

Benchmark size strongly influences sensitivity to power capping. Unlike frequency scaling, which shows high sensitivity especially for small benchmarks, power capping results in clear saturation behavior. 
Small benchmarks, which run cooler, only show performance impact at lower power caps, while larger ones feel the cap at higher values. Setting a power cap typically does not affect performance unless the cap falls below a workload-specific threshold, example, on the A100, around 150 W for small benchmarks and 250 W for larger ones. 

When comparing GPU architectures under power capping, sensitivity to power capping is limited for A100 and A40 GPUs, as benchmarks often do not reach the device's thermal design power. A100 consistently outperforms the other accelerators across all biomolecular benchmarks, reaching near-peak performance already at moderate cap levels. This early saturation suggests that the A100 is highly power-efficient, with sufficient compute and memory headroom to maintain performance even under constrained power budgets.
The A40 and L40 display intermediate behavior. Their performance scales steadily with power until mid-range caps are reached, after which additional power provides only marginal benefits. Both architectures converge to similar throughput levels across all benchmarks, reflecting comparable efficiency profiles despite differing design targets.
The L4 stands out as the most power-sensitive GPU. Performance degrades sharply under restricted caps, highlighting its dependence on higher available power to sustain throughput. Even at the highest cap levels, the L4 remains substantially behind the other GPUs, reflecting its positioning as a low-power accelerator rather than a high-performance compute device.
Power capping thus provides a more reliable and practical mechanism for energy savings, helping identify when workloads become power-sensitive without compromising performance unnecessarily.

\begin{figure}[t]
    \centering
    
    \subfloat[Pi Solver\label{fig:STREAM}]{
        \includegraphics[width=0.45\textwidth]{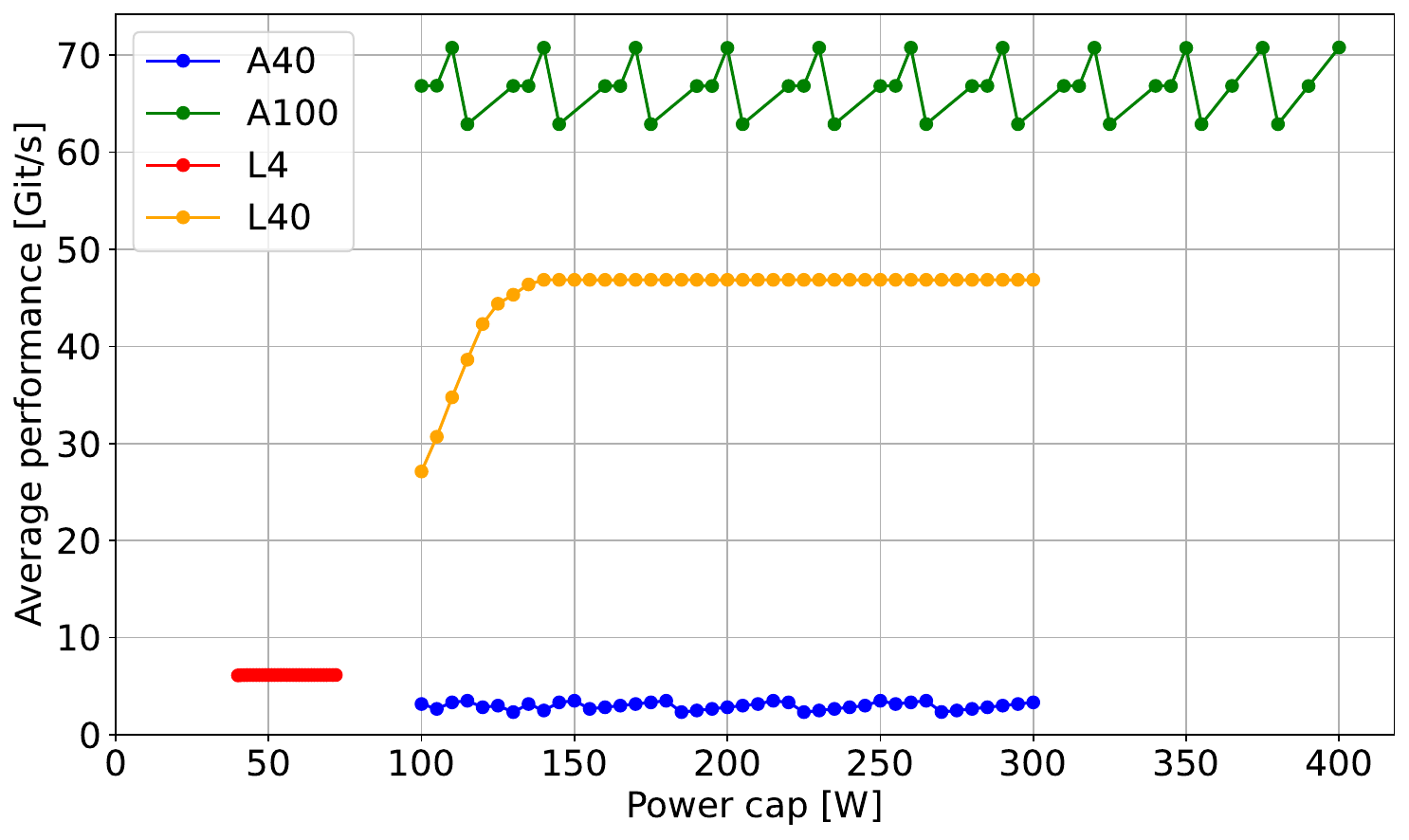}
    }
    \hfill
    \subfloat[STREAM Triad\label{fig:PISOLVER}]{
        \includegraphics[width=0.45\textwidth]{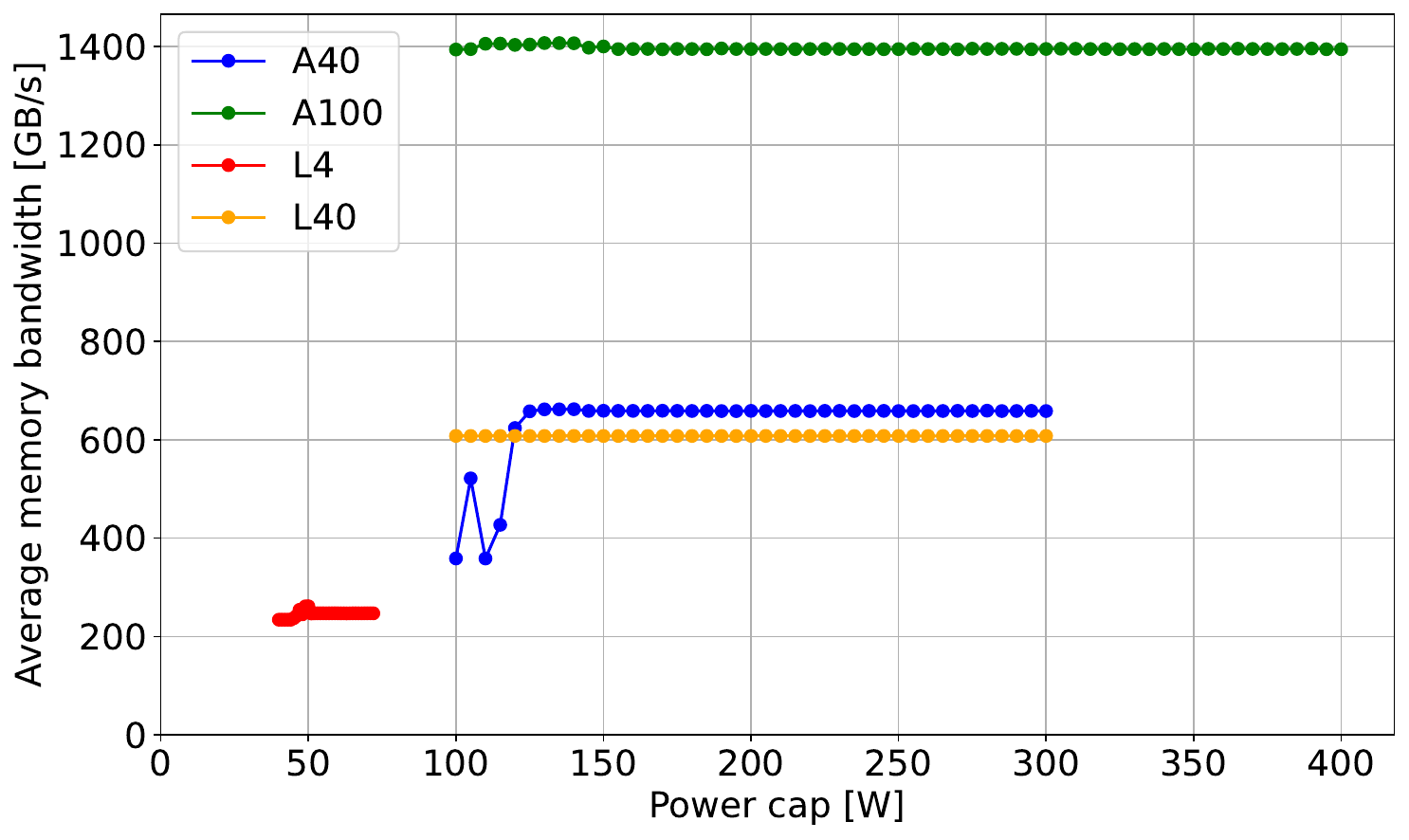}
    }
    
    \caption{Average performance as a function of power cap for various accelerators at the maximum memory frequency setting, across the Pi solver and STREAM TRIAD benchmarks.}
    \label{fig:PisolverSTREAM_perf_vs_powercap}
\end{figure}
\subsection{Pi Solver \& STREAM Triad}
Figure~\ref{fig:PisolverSTREAM_perf_vs_powercap} shows the average throughput as a function of power cap, isolating the impact of graphics frequency on the compute-bound Pi Solver and memory-bound BabelStream Triad benchmarks across the A40, A100, L4, and L40 GPUs. The Pi Solver and BabelStream benchmarks exhibit consistently low power consumption across all tested GPUs, and thus no longer represent power-intensive corner cases. Both benchmarks behave near similar under power capping: their power draw remains well below the TDP on all architectures. On the A100, BabelStream's power usage stabilizes just below 200 W, indicating that 200 W is effectively its natural power envelope, and serves as a reference point for understanding GPU power limits under memory-bound workloads.

For Pi Solver, power usage does not exceed 140 W on any device, and only the L40 shows measurable performance degradation when the power cap drops below this threshold. All other GPUs remain unaffected due to the benchmark's low compute and memory intensity. Similarly, BabelStream consistently draws less than 200,W, indicating that its memory-bound nature underutilizes available compute and bandwidth resources, especially on high-end GPUs like the A100 and L40.
On CPUs, however, BabelStream draws relatively more power than on GPUs, suggesting architectural inefficiencies such as lower memory bandwidth and reduced vectorization \cite{AfzalHW:2024,Afzal:2023:2,AfzalThesis:2015}. In contrast, Pi Solver remains cold on both platforms.
Overall, neither benchmark fully saturates both compute and memory subsystems. Their low power draw makes them largely insensitive to power capping, limiting their utility for power modeling.

\section{Conclusions and future work}\label{sec:conclude}
This work presents a comprehensive evaluation of GPU performance behavior for molecular dynamics simulations using \texttt{GROMACS}, framed within the broader context of synthetic compute- and memory-bound workloads. We analyzed four modern NVIDIA GPUs -- A40, A100, L4, and L40 -- under controlled frequency and power settings, using six biomolecular benchmarks alongside two synthetic workloads (Pi Solver and STREAM Triad).

\paragraph{Key Implications on frequency tuning}
Our results show that the sensitivity of \texttt{GROMACS} performance to GPU graphics clock frequency is highly dependent on system size and workload characteristics. Small-scale simulations benefit significantly from higher graphics clock frequencies, showing nearly linear performance improvements until saturation. In contrast, large-scale systems exhibit limited sensitivity to frequency increases, as performance becomes dominated by memory bandwidth and interconnect limitations. This mirrors the expected behavior of the Pi Solver and STREAM Triad benchmarks, which represent the compute- and memory-bound extremes, respectively.
Across architectures, the L40 consistently delivered the highest \texttt{GROMACS} throughput at peak frequency, particularly for larger systems, owing to its high bandwidth and efficient memory hierarchy. The A100 showed similar scaling on large workloads with minimal sensitivity to graphics clock frequency, indicating compute headroom and ample bandwidth. Mid-range GPUs like the A40 benefited more from frequency tuning on smaller systems, while the low-power L4 showed early saturation, suggesting a sweet spot at moderate frequencies for energy-efficient MD workloads.

\paragraph{Key implications on power capping}

Power capping results reveal that performance remains largely unaffected until power limits approach workload- and architecture-specific thresholds. High-end GPUs like the A100 and L40 maintain near-peak performance even under moderately restricted power caps, demonstrating high energy efficiency. The A100, in particular, shows early performance saturation, suggesting that it delivers excellent performance-per-watt across a broad range of power budgets.
In contrast, low-power GPUs such as the L4 are significantly more sensitive to power constraints. Their performance drops sharply under aggressive capping, limiting their scalability and making them more suitable for lightly-loaded or energy-constrained environments. Mid-range GPUs like the A40 and L40 show steady scaling under power caps, with diminishing returns as power approaches TDP.

\paragraph{Discussion and recommendations}
Together, these findings provide practical guidance for optimizing molecular dynamics workloads on heterogeneous GPU architectures. For performance-critical applications, selecting GPUs with high memory bandwidth and applying frequency tuning can yield meaningful gains for smaller benchmarks. For energy-aware computing, power capping offers a viable method to reduce consumption without sacrificing throughput—especially on high-end devices.
By combining real-world MD benchmarks with synthetic corner-case workloads, we establish a robust, reproducible framework for characterizing GPU behavior. This approach supports informed decisions in HPC system procurement, workload scheduling, and software tuning for scientific simulations at scale.

\paragraph{Future work}
Future research will explore multi-GPU scaling and the impact of CPU–GPU coordination on MD performance. Additionally, we will incorporate power modelling using synthetic workloads that operate near the ``knee'' of the Roofline model \cite{Roofline:2009} -- stressing both compute and memory -- to better characterize performance sensitivity under constrained power budgets and to enable accurate energy-efficiency modeling.


\section*{Acknowledgments}
This research is supported by EEC, a central initiative of the National High-Performance Computing at German universities (NHR), focused on enhancing energy efficiency and managing operational costs across NHR centers.
The authors gratefully acknowledge the HPC resources provided by the Erlangen National High Performance Computing Center (NHR@FAU) of the Friedrich-Alexander-Universität Erlangen-Nürnberg (FAU).
NHR funding is provided by the German Federal Ministry of Education and Research and the state governments participating on the basis of the resolutions of the GWK for the national high-performance computing at universities (www.nhr-verein.de/unsere-partner) by federal and Bavarian state authorities. 
NHR@FAU hardware is partially funded by the German Research Foundation (DFG) -- 440719683.

\bibliographystyle{splncs04}
\bibliography{references}

\end{document}